\documentclass[reprint, superscriptaddress, prb]{revtex4-1}
\usepackage[colorlinks=true,citecolor=blue,linkcolor=magenta]{hyperref}
\usepackage{lineno}
\usepackage{amsmath}
\usepackage{amsfonts}
\usepackage{stmaryrd}
\usepackage{amssymb}
\usepackage{graphicx}
\usepackage{dcolumn}
\usepackage[british]{babel}
\usepackage{bm}

\begin{document}
\title{Multiplexed control scheme for scalable quantum information processing with superconducting qubits}
\author{Pan Shi}
\thanks{These authors contributed equally to this work.}
\affiliation{Beijing Academy of Quantum Information Sciences, 100193, Beijing, China}
\affiliation{School of Physics, Wuhan University, Wuhan, 430072, China}

\author{Jiahao Yuan}
\thanks{These authors contributed equally to this work.}
\affiliation{Shenzhen Institute for Quantum Science and Engineering, Southern University of Science and Technology, Shenzhen 518055, China}
\affiliation{International Quantum Academy, Shenzhen 518048, China}
\affiliation{Guangdong Provincial Key Laboratory of Quantum Science and Engineering, Southern University of Science and Technology, Shenzhen 518055, China}
\affiliation{Department of Physics, Southern University of Science and Technology, Shenzhen 518055, China}

\author{Fei Yan}
\email{yanfei@baqis.ac.cn}
\affiliation{Beijing Academy of Quantum Information Sciences, 100193, Beijing, China}

\author{Haifeng Yu}
\affiliation{Beijing Academy of Quantum Information Sciences, 100193, Beijing, China}

\begin{abstract}
\textbf{The advancement of scalable quantum information processing relies on the accurate and parallel manipulation of a vast number of qubits, potentially reaching into the millions. Superconducting qubits, traditionally controlled through individual circuitry, currently face a formidable scalability challenge due to the excessive use of wires. This challenge is nearing a critical point where it might soon surpass the capacities of on-chip routing, I/O packaging, testing platforms, and economically feasible solutions. Here we introduce a multiplexed control scheme that efficiently utilizes shared control lines for operating multiple qubits and couplers. By integrating quantum hardware-software co-design, our approach utilizes advanced techniques like frequency multiplexing and individual tuning. This enables simultaneous and independent execution of single- and two-qubit gates with significantly simplified wiring. This scheme has the potential to diminish the number of control lines by one to two orders of magnitude in the near future, thereby substantially enhancing the scalability of superconducting quantum processors.
}
\end{abstract}

\maketitle


In the rapidly advancing field of quantum computing, superconducting quantum circuits have emerged as a promising candidate for scalable quantum information processing. State-of-the-art processors have now reached scales of approximately 100 qubits, as evidenced by recent progress~\cite{acharya_suppressing_2023, cao2023generation, xu2023digital, kim_evidence_2023}. Despite their potential, a paramount challenge hindering the development of large-scale quantum processors is the existing control infrastructure. With each qubit requiring dedicated control lines for gate operations, the current control architecture would become increasingly untenable as the system expands. This is not just a matter of physical wiring; the complexity extends to signal routing at the quantum chip level, space and cooling power limitations within cryogenic systems, and the introduction of additional noise channels, all of which exacerbate the scalability issue. Moreover, the increasing number of control electronics and components strains the economic feasibility of scaling these systems. 
While various strategies are being explored to mitigate these issues~\cite{asaad_independent_2016, PhysRevApplied.19.054050, bejanin2022quantum, lecocq_control_2021}, they often resort to time-division multiplexing, contradicting the goal of parallel operations, or offer only partially multiplexed signal delivery. There exists a notable gap in comprehensive solutions that enable the simultaneous orchestration of both single- and two-qubit gates while significantly reducing the wire count at the chip level.

In this manuscript, we introduce a novel multiplexed control scheme intended to operate multiple qubits or couplers in parallel by integrating quantum hardware-software co-design. 
The scheme is tailored for the implementation of error-correcting codes as well as general circuits on a state-of-the-art architecture built with superconducting transmon qubits~\cite{koch2007charge} and tunable transmon couplers~\cite{yan2018tunable}. 
By leveraging frequency multiplexing on shared microwave lines and integrating a pulse-shape-invariant compiling method, our approach can effectively suppress leakage and coherent errors while executing simultaneous arbitrary single-qubit gates.
Furthermore, to facilitate simultaneous controlled-Z gates across arbitrary qubit pairs, we utilize shared flux bias lines and microwave tuning technique, in conjunction with an efficient circuit decomposition involving $\sqrt{\mathrm{CZ}}$ gates. 
We also present a variation of the scheme based on fluxonium qubits~\cite{manucharyan2009fluxonium,nguyen2019high} and fixed-frequency transmon couplers, enabling all-microwave control. Our proposed scheme promises a reduction in wire requirements by one to two orders of magnitude in the near future, suggesting a commensurate expansion in the feasible system size for superconducting quantum processors.

The paper is structured as follows: we first outline the general concept for realizing multiplexed control in an architecture based on fixed-frequency transmon qubits and tunable transmon couplers. We briefly describe how precise and parallel single- and two-qubit gates are realized in this architecture. We then discuss in detail the leakage processes during single-qubit gate operations and elucidate the individual tuning technique for two-qubit controlled phase gates. Lastly, we discuss several pertinent aspects related to the scalability of our proposed scheme.

\begin{figure*}[t]
\begin{center}
	\includegraphics[scale=1.0]{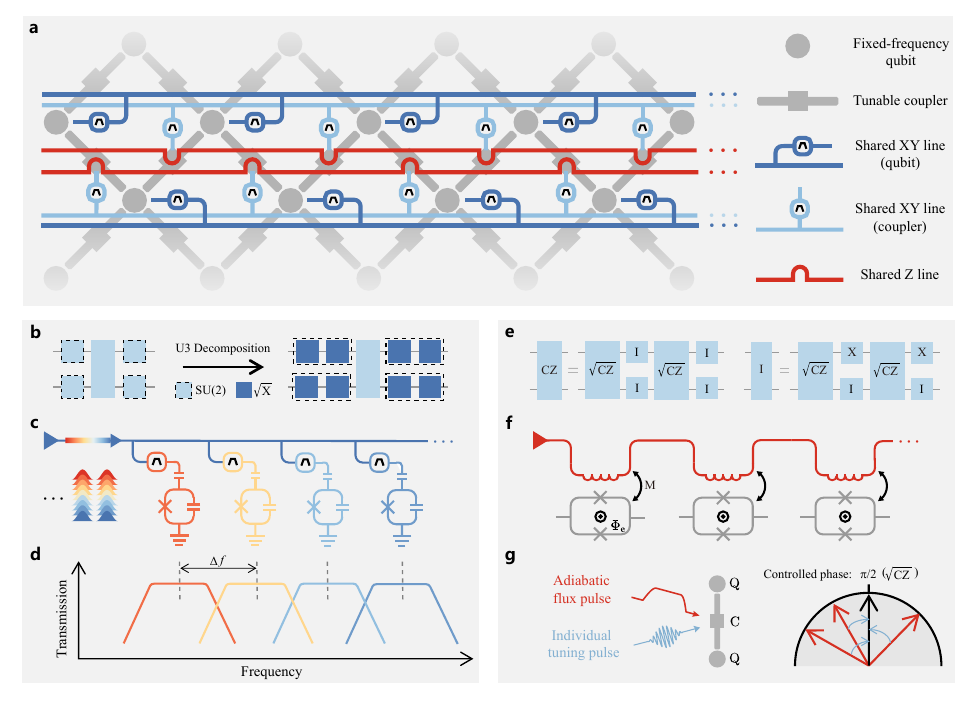}
	\caption{
    \label{fig1}
    {\bf Multiplexed control scheme using shared control lines.}
    {\bf a,} Schematic layout of the drive lines for controlling a two-dimensional array of fixed-frequency qubits and tunable couplers.
    The dark (light) blue lines denote shared charge drive (XY) lines for simultaneously delivering microwave signals to multiple qubits (couplers), with dedicated bandpass filters used for signal isolation. The red lines denote shared flux drive (Z) lines, delivering flux bias signals to multiple couplers at once. 
    {\bf b,} Compiling scheme for single-qubit gates. During single-qubit gate cycles, any SU(2) operation -- including Identity -- is decomposed, according to the U3 decomposition, into two $\sqrt{\mathrm{X}}$ gates and three virtual-Z gates. Consequently, the single-qubit gate cycles are filled with pairs of $\pi/2$ pulses for all qubits.
    {\bf c,} Circuit diagram of a shared XY line which functions as a distributed multiplexer consisting of a series of bandpass filters, delivering superimposed microwave pulses of different frequencies (indicated by different colors) to the corresponding qubits.
    {\bf d,} Transmission characteristics of the multiplexer. The filter passbands are evenly spaced with a spacing of $\Delta f$; the bandwidth is also approximately $\Delta f$.
    {\bf e,} Decomposition of the controlled-Z (CZ) gate and the Identity gate using two $\sqrt{\mathrm{CZ}}$ gates and single-qubit operations. Therefore, $\mathrm{CZ}$ operations on an arbitrary selection of qubit pairs can be realized using simultaneous $\sqrt{\mathrm{CZ}}$ gates.
    {\bf f,} Circuit diagram of a shared Z line which inductively couples to the SQUID loops of multiple couplers, applying a near-uniform magnetic flux ($\Phi_\mathrm{e}$) across these SQUIDs.
    {\bf g,} Left: Conceptual representation of individually tuning a controlled-phase gate using a common adiabatic flux pulse (red pulse) alongside an individual microwave tuning pulse (blue pulse).
    Right: The common flux pulse facilitates a controlled phase gate close to a $\sqrt{\mathrm{CZ}}$ gate, with individual tuning pulses adjusting for uniformity across qubit pairs.
    }
\end{center}
\end{figure*}

\section*{The general architecture}

Figure \ref{fig1}{\bf a} shows a two-dimensional array of superconducting transmon qubits arranged in a square lattice. The configuration aligns with the surface code, which is naturally suitable for superconducting qubits~\cite{PhysRevA.86.032324}. 
Here, we employ tunable transmon couplers between each pair of neighboring qubits for independent modulation of their couplings, a strategy proven to effectively mitigate crosstalk and facilitate high-fidelity parallel quantum gates in recent years.
In the traditional control infrastructure, each qubit requires a dedicated charge drive line; each tunable coupler requires a dedicated flux bias line.

Central to our design is the simplified arrangement of control lines.
The qubits on each row share a common charge drive line, which interacts, through capacitive coupling, with the qubits' Pauli-X and Pauli-Y terms, thus named the XY line. The couplers in the same row are connected to two flux drive lines, which, through inductive coupling, modulate their frequencies by interacting with the couplers' Pauli-Z terms, hence referred to as the Z line. These two Z lines are arranged in an interleaved fashion, considering the operational requirement that couplers connected to the same qubit are activated sequentially, such as during parity check cycles in the surface code. 
Additionally, there are shared XY lines that connect to the couplers; we shall discuss their usage later.

This architecture forms the backbone of our multiplexed control strategy, enabling high-fidelity arbitrary single-qubit gates in parallel to all the qubits connected to the same XY line. Moreover, it facilitates high-fidelity universal two-qubit gates in parallel to an arbitrary subset of qubit pairs connected to the same Z line. These capabilities are crucial for executing most quantum circuits or algorithms, including the surface code.

An important consideration in our design is the inherent variability in superconducting qubits' frequencies due to fabrication challenges. 
Recent advancements in fabrication technology, however, have led to remarkable improvements in frequency targeting~\cite{Kreikebaum_2020,10.1063/5.0037093}. The best-reported standard variation in qubit frequencies is approximately 10 MHz~\cite{doi:10.1126/sciadv.abi6690}.
This improvement is instrumental in our scheme, as it directly impacts the efficiency and scalability of the multiplexed control system.

\textbf{Single-qubit gates using a shared microwave line.}
In our architecture, we employ a novel approach for the control of single-qubit gates using a shared microwave line. The process begins with reorganizing the quantum circuit into interleaved layers of single- and two-qubit gates. Subsequently, we deploy the U3 decomposition method to compile single-qubit gates, including the Identity operation, into two $\sqrt{X}$ gates and three virtual-Z gates, as depicted in Fig.~\ref{fig1}{\bf b}~\cite{mckay2017efficient}. This compilation technique, known as the isomorphic waveform method~\cite{han2023multilevel}, ensures that all qubits undergo a uniform gate cycle, consisting of pairs of $\pi/2$ pulses.

To enable parallel signal delivery, we apply the technique of frequency multiplexing, widely used in classical communications. This involves designing the qubits connected by the same XY line to have uniformly spaced frequencies, with an interval of $\Delta f=50$ MHz, for example. The $\pi/2$ pulses at these distinct frequencies are then superimposed in the time domain and transmitted through the shared XY line. At each qubit’s branch, a bandpass filter is utilized, aligned with the qubit’s specific frequency, to prevent out-of-band frequencies from infiltrating. This setup, illustrated in Fig.~\ref{fig1}{\bf c} and Fig.~\ref{fig1}{\bf d}, effectively acts as a multiplexer, delivering the correct frequency to each qubit.

Achieving high fidelity across all qubits in this shared line system presents a significant challenge, particularly due to the presence of signals at various frequencies and their inherent randomness. A primary concern is the occurrence of leakage transitions—those other than the desired $|0\rangle \leftrightarrow |1\rangle$ transition. These unwanted transitions can shift the qubit state outside the computational subspace, posing a challenge to correct. Our strategy employs bandpass filters to significantly reduce the likelihood of these leakage transitions by attenuating signals that could potentially induce them. We shall discuss leakage errors in detail later. 

Another critical error source is the Stark effect, induced by off-resonant drives. Our approach to mitigating this involves the simultaneous calibration of the superposed $\pi/2$ pulses in terms of both amplitude and frequency, a process facilitated by the isomorphic waveform method. This calibration ensures that the pulses maintain a consistent shape and, consequently, a predictable influence on crosstalk. By combining this circuit compilation strategy with the multiplexer design, we can execute high-fidelity arbitrary single-qubit gates in parallel, effectively addressing both leakage and Stark effect-induced errors. 

\textbf{Two-qubit gates using a shared flux bias line.}
In our quantum processor architecture, a basic building block is the two-qubit controlled-Z or CZ (controlled phase: $\pi$) gate, which is equivalent to the universal CNOT gate up to single-qubit operations. Central to our scheme is the efficient execution of the CZ gate across select qubit pairs, facilitated by a shared flux bias line.

We introduce a circuit decomposition scheme, as shown in Fig.~\ref{fig1}{\bf e}, that transforms both the CZ and the Identity operations into a combination of two $\sqrt{\mathrm{CZ}}$ (controlled phase: $\pi/2$) gates and single-qubit gates.  This approach allows for the parallel implementation of CZ gates on any chosen pair of qubits, provided that parallel $\sqrt{\mathrm{CZ}}$ gates can be realized. The capability to perform arbitrary single-qubit operations in parallel, as previously discussed, further complements this strategy.

In our design, the shared Z line is inductively coupled to the SQUID loops of multiple tunable couplers, as depicted in Fig.~\ref{fig1}{\bf f}. This design aims for a relatively uniform mutual inductance and consistent flux amplitudes from a common pulse sent through the shared Z line. Employing a coupler-assisted adiabatic controlled-phase gate scheme~\cite{xu2020high, PhysRevLett.125.240502}, which necessitates only modulation of the coupler frequency or flux, we are able to induce approximate $\sqrt{\mathrm{CZ}}$ gates across all qubit pairs. However, the individual controlled phases generated by this common flux pulse may vary among different qubit pairs.

To address this variance and achieve the desired phase values, we introduce concurrent microwave pulses directed to the couplers through another shared XY line, also a multiplexer with filter passbands aligned with the coupler frequencies. These pulses drive frequency-selective transitions in the couplers, dependent on the qubit states, allowing for the generation of controlled phases that can be individually adjusted. We shall discuss the mechanism in detail later.

Essentially, our approach to executing parallel CZ gates integrates a common controlled-phase gate (via the shared Z line) with individual phase tuning (using a shared XY line). This strategy, in conjunction with our circuit compiling methods, significantly reduces the requisite number of control lines, thereby enhancing the scalability and efficiency of our quantum processor.

\begin{figure}[t]
\begin{center}
	\includegraphics[scale=0.90]{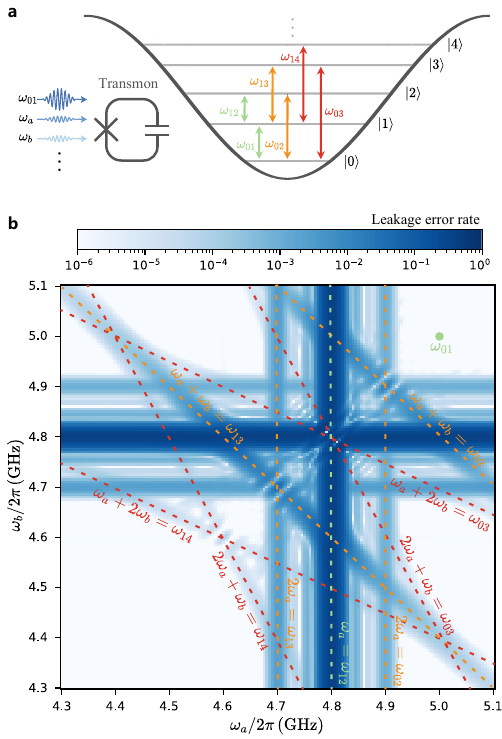}
	\caption{
    \label{fig2}
    {\bf Leakage error analysis for transmon qubits.}
    {\bf a,} Center: Energy level diagram of a transmon qubit. The lowest two levels, $|0\rangle$ and $|1\rangle$, are the computational states. Higher energy levels are considered as non-computational or leakage states. Noteworthy transitions include single-photon $|0\rangle \leftrightarrow |1\rangle$ and $|1\rangle \leftrightarrow |2\rangle$, two-photon $|0\rangle \leftrightarrow |2\rangle$ and $|1\rangle \leftrightarrow |3\rangle$, and three-photon $|0\rangle \leftrightarrow |3\rangle$ and $|1\rangle \leftrightarrow |4\rangle$. 
    Left: The transmon qubit is exposed not only to the main signal at the target frequency $\omega_{01}$ but also to signals for nearby qubits ($\omega_a$, $\omega_b$, etc.), although the amplitudes may be reduced due to filtering.
    {\bf b,} Numerically simulated total leakage error rate versus varying frequencies of two spurious signals ($\omega_a$ and $\omega_b$) assuming no attenuation with a 50-ns cosine-shaped $\pi/2$ pulse. The qubit frequency is set to $\omega_{01}/2\pi=5$GHz, indicated by the green dot. The results unveil stripe patterns that correspond to the different transitions shown in {\bf a}.
    }
\end{center}
\end{figure}

\section*{Leakage errors of single-qubit gates}

In our scheme, effectively addressing leakage errors in single-qubit gates is crucial, particularly for transmon qubits. The transmon qubit, essentially comprising a Josephson junction and a shunt capacitor, is favored for its simplicity and high reproducibility. Despite these advantages, the transmon's level structure, akin to a weakly anharmonic oscillator as shown in Fig.~\ref{fig2}{\bf a}, presents inherent challenges. The low anharmonicity, or limited variation in inter-level spacing, is a key reason behind leakage transitions, where qubits inadvertently transition out of the computational basis due to closely spaced signal frequencies.

Dominant leakage transitions in transmon qubits include not only the single-photon 1-2 transition but also higher-order processes like two-photon and three-photon transitions. To understand and quantify these transitions under our multiplexed control protocol, we conducted simulations that investigate the influence of spurious signals on leakage error rates. These simulations, incorporating two additional spurious signals, $\omega_a$ and $\omega_b$, alongside the primary $\pi/2$ pulse—all with equivalent amplitudes—were set to typical transmon qubit configurations: a 0-1 transition frequency at 5 GHz and an anharmonicity of -200 MHz.

The simulation results, as shown in Fig.~\ref{fig2}{\bf b}, exhibit distinctive stripe patterns that correlate with various types of leakage transitions. Notably, there is a significant increase in error rates, between 0.1 and 1, when spurious frequencies closely align with the sensitive 1-2 transition frequency at 4.8 GHz. In contrast, the two-photon transitions, such as $|0\rangle \leftrightarrow |2\rangle$ and $|1\rangle \leftrightarrow |3\rangle$, exhibit lower error rates on the order of $10^{-3}$ due to their second-order nature. Additionally, three-photon transitions display even lower error rates, around $10^{-5}$.
Note that various frequency combinations can induce specific transitions, such as $|0\rangle \leftrightarrow |2\rangle$, which include combinations like $2\omega_a=\omega_{02}$, $2\omega_b=\omega_{02}$, and $\omega_a+\omega_b=\omega_{02}$.

To mitigate these leakage errors, our design employs bandpass filters, which are highly effective in attenuating signals that deviate from the passband. For instance, if a bandpass filter centered at 5 GHz can attenuate signals at 4.8 GHz and 4.9 GHz by at least 40 dB and 10 dB respectively, we can suppress both single-photon and two-photon leakage errors to a rate of about $10^{-5}$. It's crucial to note that while the transition rate for single-photon processes scales linearly with signal power, it scales quadratically for two-photon processes.
These insights are instrumental in shaping the design goals for the multiplexer, ensuring it effectively minimizes leakage errors.

\begin{figure*}[t]
\begin{center}
	\includegraphics[width=1\textwidth]{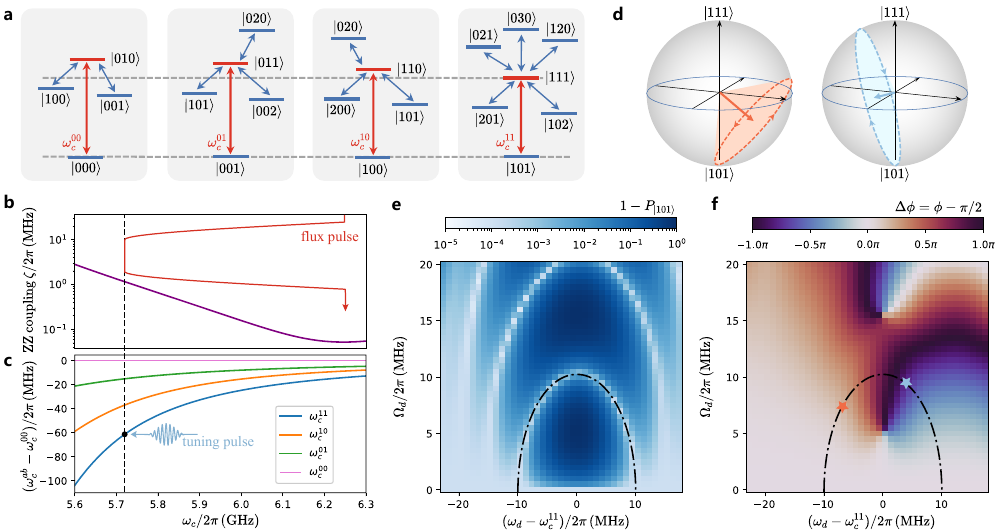}
	\caption{
    \label{fig3}
    {\bf Individual tuning of controlled phase for transmon qubits.}
    {\bf a,} Energy level diagrams showing different coupler transitions $|a0b\rangle \leftrightarrow |a1b\rangle$ ($a,b=0,1$) for the four computational states, illustrating distinct frequency shifts due to level repulsion.
    {\bf b-c,} The longitudinal ZZ interaction strength $\zeta$ and the coupler's transition frequency relative to $\omega_c^{00}$ for the four computational bases (i.e., $\omega_c^{ab}-\omega_c^{00}$) as a function of the coupler's bare frequency $\omega_c$. 
    In the simulation, the qubit frequencies are 5.3 GHz and 5.0 GHz, respectively; the anharmonicities are -200 MHz for both qubits and coupler; the qubit-coupler coupling strengths are 100 MHz; the qubit-qubit coupling strength is 10 MHz.
    According to the scheme described in Fig.~\ref{fig1}{\bf g}, a flux pulse applied to the coupler (red pulse) shifts the system to a bias (dashed line) where the ZZ interaction is enhanced appreciably, performing a controlled-phase gate.
    At this bias, the discrepancy of the coupler frequency is drastically increased. One can apply a microwave pulse near the frequency $\omega_c^{11}$ (blue pulse) on top of the flux pulse via the shared coupler-XY line for selective driving.
    {\bf d,} Bloch sphere representations of the trajectories during the selective coupler transition $|101\rangle \leftrightarrow |111\rangle$, showing geometric phase accumulation. 
    The example on the left shows the case with a smaller solid angle (greater detuning) compared to the case on the right.
    {\bf e-f,} Simulated non-$|101\rangle$ state population and conditional phase in response to the tuning pulse frequency $\omega_d$ and amplitude $\Omega_d$. In the simulation, we use a 200-ns cosine-shaped pulse on top of a 200-ns flattop flux pulse with an additional 20-ns rise and fall.
    The dot-dashed line indicates the drive parameters that cause minimal population out of the $|111\rangle$ state. Along the same curve in {\bf f}, a full range of conditional phases can be achieved. The stars mark the two corresponding cases in {\bf d}.
    }
\end{center}
\end{figure*}

\section*{Individual tuning of two-qubit gates}

As previously introduced, in our scheme, precise $\sqrt{\mathrm{CZ}}$ operation for each qubit pair is accomplished through a coupler-assisted controlled-phase gate induced by a common coupler flux pulse, and individual tuning of the controlled phase by applying a microwave pulse to the coupler. 
To facilitate simultaneous operations, we ensure that, akin to the qubits, the coupler frequencies are uniformly spaced. This allows for the superposition of microwave pulses, which can then be directed to the appropriate couplers using a multiplexer.

As depicted in Fig.~\ref{fig3}{\bf a}, the couplers' excited states interact with a qubit-state-dependent environment, exhibiting distinct sets of neighboring energy levels. Due to level repulsion, this interaction causes the separation of the couplers' transition frequencies for different qubits states ($\omega_c^{ab}$, where $a,b=0,1$ represent qubit states).
The extent of frequency separation depends on both the interaction strengths and detunings between energy levels. It becomes more pronounced when the coupler is flux-biased towards lower frequencies, closer to the qubit frequencies, thereby intensifying the level repulsion and resulting in a frequency separation of tens of megahertz, as shown in Fig.~\ref{fig3}{\bf b} and \ref{fig3}{\bf c}. 

These coupler-assisted interactions, interactions that can be adjusted by the working point of the coupler~\cite{chu_coupler-assisted_2021}, enable highly selective driving of a specific transition, like $|101\rangle \leftrightarrow |111\rangle$ at frequency $\omega_c^{11}$.
It can be utilized in realizing two-qubit entangling gates, including resonant microwave controlled-phase gates, as evidenced in prior works~\cite{PhysRevX.11.021026, PhysRevX.13.031035}. For instance, driving the $|101\rangle \leftrightarrow |111\rangle$ transition with a constant detuning and amplitude causes the system to trace a small circle on the Bloch sphere and eventually return to the original $|101\rangle$ state, as illustrated in Fig.~\ref{fig3}{\bf d}. Although the population of the states remains unchanged after this process, a geometric phase accumulates on the $|101\rangle$ state. This phase, equivalent to the solid angle enclosed by the trajectory on the Bloch sphere, effectively manifests as a controlled phase. 

This geometric phase can be conveniently tuned by joint adjustment of the drive frequency $\omega_d$ and the amplitude $\Omega_d$, assuming a fixed pulse length. Our simulations, presented in Fig.\ref{fig3}{\bf e} and \ref{fig3}{\bf f}, demonstrate how varying these parameters impacts the final state population and the controlled phase. The parameters that ensure the system returns to the $|101\rangle$ state are indicated by a dot-dashed curve in Fig.\ref{fig3}{\bf e}. The effective rotation rate, calculated as $\sqrt{\Omega_d^2 + (\omega_d - \omega_c^{11})^2}$, remains constant along this curve due to the fixed total rotation angle of $2\pi$. As shown in Fig.\ref{fig3}{\bf f}, this method allows for a wide range of conditional phases, spanning from $-\pi$ to $\pi$, thus highlighting the adaptability of this technique. However, it is noteworthy that higher drive amplitudes may affect drive selectivity, potentially impacting performance.
In practice, using a shaped waveform can further suppress state leakage via other coupler transitions. The additional phases accumulated on the other three computational states due to the far-detuned drive can be accounted for during the calibration of the controlled and local phases.

\begin{figure}[t]
\begin{center}
        \includegraphics[width=0.5\textwidth]{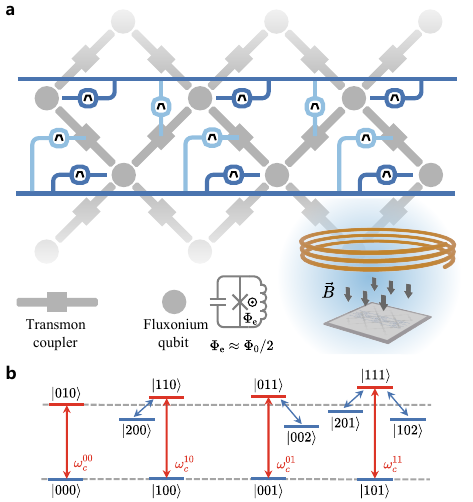}
	\caption{
    \label{fig4}
    {\bf All-microwave scheme with fluxonium qubits.}
    {\bf a,} Layout of the shared drive lines for controlling a two-dimensional array of fluxonium qubits in conjunction with fixed-frequency transmon couplers.
    Here, to further simplify wiring, a multiplexer is used to deliver control signals for both qubits and couplers, as they are designed to operate at distinct frequencies. 
    Shown in the lower right inset is the proposed global uniform magnetic field for biasing all fluxonium qubits near half a flux quantum.
    {\bf b,} Energy level diagrams of the four different coupler transitions $|a0b\rangle \leftrightarrow |a1b\rangle$ ($a,b=0,1$) corresponding to the four computational states for fluxonium qubits, emphasizing significant frequency shifts due to strong qubit-coupler coupling, enabling selective driving for two-qubit gate operations.
    }
\end{center}
\end{figure}

\section*{All-microwave scheme for fluxoniums}

We next introduce an alternative variant of our multiplexed control scheme, tailored for fluxonium qubits—a promising modality in superconducting quantum information processing. Fluxonium qubits offer greater flexibility in energy level structuring compared to transmon qubits. Notably, they can achieve large anharmonicities, with the 1-2 transition frequency exceeding the 0-1 transition frequency by typically a few gigahertz, when biased at half a flux quantum, known as the 'sweet spot' where coherence times are often substantially extended.

Exploiting these advantages, we propose an all-microwave control scheme complemented by a global DC flux bias, ideally suited for fluxonium qubits (as shown in Fig.~\ref{fig4}{\bf a}). This scheme operates at the sweet spot, leveraging the benefits of both extended coherence times and large anharmonicity. Unlike the localized bias lines, one can uniformly bias all the qubits near their sweet spots by applying a global magnetic field, for instance, using a large coil. This approach drastically simplifies the wiring efforts. However, it may be a challenge to create a highly homogeneous magnetic field.

For single-qubit gate operations, we employ the frequency-multiplexing method, similar to the one used for transmon qubits, via a shared XY line. The two-qubit gate control scheme is similar to the geometric phase gate approach used for transmon qubits, with an important distinction: fluxonium qubits, due to their larger separation of higher energy levels, allow for stronger coupling—up to about 500 MHz—with the transmon coupler while maintaining minimal static ZZ coupling (less than 10 kHz)~\cite{PhysRevX.13.031035}. Additionally, if the fluxoniums' 1-2 transition frequencies are near the transmon coupler's 0-1 transition frequency, it naturally results in a substantial frequency separation, possibly over 100 MHz, among the coupler frequencies corresponding to different computational states (as shown in Fig.~\ref{fig4}{\bf b}). This feature enables selective driving without the need for a tunable coupler or any Z control line.

Similarly, these fixed-frequency transmon couplers can be uniformly spaced in frequency (e.g., at intervals of about 100 MHz) and controlled using a shared XY line equipped with dedicated bandpass filters. For processing general circuits, we apply the previously introduced U3 decomposition and $\sqrt{\mathrm{CZ}}$ decomposition techniques. This approach allows for simultaneous yet independent single- and two-qubit gates, aligning with the requirements of various quantum algorithms and error-correcting codes.

\section*{Discussions}

Lastly, we discuss several important considerations and long-term projections regarding the scalability and practicality of our multiplexed control scheme. 

\textbf{Multiplicity.}
For the shared Z line, there is no fundamental limit to the number of couplers that can be connected to a single transmission line. However, for the shared XY line, the multiplicity $M$ is determined by the total frequency range $W$ available for housing the qubits, divided by their interval $\Delta f$, thus $M=W/\Delta f$. The primary limitation here is the random variation in qubit frequency. While global common-mode frequency deviation is a common scenario, it does not significantly impact our scheme. Post-fabrication design adjustments of the multiplexer chip or fabricating a series of qubit chips can effectively manage this variability. Here, the variation in the frequency interval $\Delta f$ is the most critical property. By limiting the differential-mode variation in qubit frequency to below 10 MHz, a multiplicity of several hundred is achievable within a few gigahertz spectral range. This represents a substantial reduction in the number of wires required, potentially downsizing the number of necessary cryogenic cables, microwave components, and signal generators, thus making the system more cost-effective and scalable.

\textbf{Pulse duration.}
A bandwidth limit of 10 MHz necessitates even narrower spectral widths for single-qubit control pulses, resulting in pulse widths of a few hundred nanoseconds. While longer pulses might lead to increased decoherence errors, there are compelling reasons to extend pulse durations in the long term. Longer pulses with lower power help minimize control errors, including leakage transitions and Stark effects. Future advancements in achieving longer intrinsic coherence times -- limited only by local noise sources -- will necessitate reducing the coupling strength to the external control circuitry, thereby slowing down gate operations. Fortunately, the relaxation time due to external coupling scales as $T_1 \propto 1/g_\mathrm{e}^2$, and gate time as $t_\mathrm{gate} \propto 1/g_\mathrm{e}$, indicating that reduced coupling strength will beneficially impact gate error rates $\epsilon \sim t_\mathrm{gate}/T_1 \propto g_\mathrm{e}$.

\textbf{Heat load.}
Primary contributors to heat load in a cryogenic system are heat conduction via cables (passive) and signal dissipation at the attenuators (active). Presently, a standard dilution refrigerator can handle about 1000 cables, potentially more with superconducting cables. In traditional control schemes, the active dissipation rate on a single attenuator is proportional to the average power ($P\propto\Omega_d^2$) of a single pulse; thus, the total rate is $N \times P$ for $N$ qubits with dedicated control lines. Utilizing our multiplexed scheme increases the power dissipated on a single attenuator by a factor of $M$, but extending the gate time, hence reduced drive amplitude $\Omega_d$, can help offset this. There are alternative solutions that can be helpful in this respect, such as nonreciprocal control circuits~\cite{kono_breaking_2020} and the development of cryogenic electronics~\cite{acharya_multiplexed_2023}.

In an optimistic scenario with a qubit lifetime $T_1=10$ ms, a gate error rate of about $10^{-5}$ is feasible using a 200-ns gate time for both single- and two-qubit gates. The active power needed for such control, while maintaining appropriate coupling to the environment, is approximately $P\approx \frac{\sqrt{\pi}}{6}\hbar \omega_{q} T_1 \Omega_d^2\approx -90$ dBm~\cite{krinner2019engineering}, where $\omega_q/2\pi\approx 5$ GHz and the Rabi frequency $\Omega_d/2\pi\approx 1.6$ MHz corresponding to a 200-ns $\sqrt{\mathrm{X}}$ gate. It is possible to accommodate $10^5$ superconducting qubits with 1000 cables and a multiplicity of 100 (with $\Delta f=10$ MHz) in a single cryogenic system.

\textbf{Multiplexer.}
Designing an on-chip multiplexer that meets stringent requirements, including frequency, bandwidth, and isolation, remains a technical challenge, particularly with size constraints~\cite{shang2013novel}. To further reduce leakage errors, filters with higher orders are needed, which implies increased circuit segments and a larger footprint. However, the high isolation of superconducting filters~\cite{hong2004microstrip} and potential miniaturization using superconducting metamaterial structures~\cite{mirhosseini2018superconducting} offer promising solutions.

In summary, our research introduces a novel multiplexed control scheme for scalable quantum information processing with superconducting qubits, aiming to drastically reduce control line requirements. By integrating quantum hardware and software co-design, we employ shared microwave lines for efficient parallel control of both single- and two-qubit gates. Our work provides a viable solution for building larger, more complex quantum information processing systems, facilitating future developments of quantum information technologies.




\bigskip
\bibliographystyle{naturemag}
\bibliography{bibliography}

\begin{thebibliography}{10}
\expandafter\ifx\csname url\endcsname\relax
  \def\url#1{\texttt{#1}}\fi
\expandafter\ifx\csname urlprefix\endcsname\relax\def\urlprefix{URL }\fi
\providecommand{\bibinfo}[2]{#2}
\providecommand{\eprint}[2][]{\url{#2}}

\bibitem{acharya_suppressing_2023}
\bibinfo{author}{Acharya, R.} \emph{et~al.}
\newblock \bibinfo{title}{Suppressing quantum errors by scaling a surface code
  logical qubit}.
\newblock \emph{\bibinfo{journal}{Nature}} \textbf{\bibinfo{volume}{614}},
  \bibinfo{pages}{676--681} (\bibinfo{year}{2023}).
\newblock \urlprefix\url{https://doi.org/10.1038/s41586-022-05434-1}.

\bibitem{cao2023generation}
\bibinfo{author}{Cao, S.} \emph{et~al.}
\newblock \bibinfo{title}{Generation of genuine entanglement up to 51
  superconducting qubits}.
\newblock \emph{\bibinfo{journal}{Nature}} \textbf{\bibinfo{volume}{619}},
  \bibinfo{pages}{738--742} (\bibinfo{year}{2023}).

\bibitem{xu2023digital}
\bibinfo{author}{Xu, S.} \emph{et~al.}
\newblock \bibinfo{title}{Digital simulation of projective non-abelian anyons
  with 68 superconducting qubits}.
\newblock \emph{\bibinfo{journal}{Chinese Physics Letters}}
  (\bibinfo{year}{2023}).

\bibitem{kim_evidence_2023}
\bibinfo{author}{Kim, Y.} \emph{et~al.}
\newblock \bibinfo{title}{Evidence for the utility of quantum computing before
  fault tolerance}.
\newblock \emph{\bibinfo{journal}{Nature}} \textbf{\bibinfo{volume}{618}},
  \bibinfo{pages}{500--505} (\bibinfo{year}{2023}).
\newblock \urlprefix\url{https://doi.org/10.1038/s41586-023-06096-3}.

\bibitem{asaad_independent_2016}
\bibinfo{author}{Asaad, S.} \emph{et~al.}
\newblock \bibinfo{title}{Independent, extensible control of same-frequency
  superconducting qubits by selective broadcasting}.
\newblock \emph{\bibinfo{journal}{npj Quantum Information}}
  \textbf{\bibinfo{volume}{2}}, \bibinfo{pages}{16029} (\bibinfo{year}{2016}).
\newblock \urlprefix\url{https://doi.org/10.1038/npjqi.2016.29}.

\bibitem{PhysRevApplied.19.054050}
\bibinfo{author}{Zhao, P.} \emph{et~al.}
\newblock \bibinfo{title}{Baseband control of superconducting qubits with
  shared microwave drives}.
\newblock \emph{\bibinfo{journal}{Phys. Rev. Appl.}}
  \textbf{\bibinfo{volume}{19}}, \bibinfo{pages}{054050}
  (\bibinfo{year}{2023}).
\newblock
  \urlprefix\url{https://link.aps.org/doi/10.1103/PhysRevApplied.19.054050}.

\bibitem{bejanin2022quantum}
\bibinfo{author}{B{\'e}janin, J.}, \bibinfo{author}{Earnest, C.} \&
  \bibinfo{author}{Mariantoni, M.}
\newblock \bibinfo{title}{The quantum socket and demuxyz-based gates with
  superconducting qubits}.
\newblock \emph{\bibinfo{journal}{arXiv preprint arXiv:2211.00143}}
  (\bibinfo{year}{2022}).

\bibitem{lecocq_control_2021}
\bibinfo{author}{Lecocq, F.} \emph{et~al.}
\newblock \bibinfo{title}{Control and readout of a superconducting qubit using
  a photonic link}.
\newblock \emph{\bibinfo{journal}{Nature}} \textbf{\bibinfo{volume}{591}},
  \bibinfo{pages}{575--579} (\bibinfo{year}{2021}).
\newblock \urlprefix\url{https://doi.org/10.1038/s41586-021-03268-x}.

\bibitem{koch2007charge}
\bibinfo{author}{Koch, J.} \emph{et~al.}
\newblock \bibinfo{title}{Charge-insensitive qubit design derived from the
  cooper pair box}.
\newblock \emph{\bibinfo{journal}{Phys. Rev. A}} \textbf{\bibinfo{volume}{76}},
  \bibinfo{pages}{042319} (\bibinfo{year}{2007}).
\newblock \urlprefix\url{https://link.aps.org/doi/10.1103/PhysRevA.76.042319}.

\bibitem{yan2018tunable}
\bibinfo{author}{Yan, F.} \emph{et~al.}
\newblock \bibinfo{title}{Tunable coupling scheme for implementing
  high-fidelity two-qubit gates}.
\newblock \emph{\bibinfo{journal}{Phys. Rev. Appl.}}
  \textbf{\bibinfo{volume}{10}}, \bibinfo{pages}{054062}
  (\bibinfo{year}{2018}).
\newblock
  \urlprefix\url{https://link.aps.org/doi/10.1103/PhysRevApplied.10.054062}.

\bibitem{manucharyan2009fluxonium}
\bibinfo{author}{Manucharyan, V.~E.}, \bibinfo{author}{Koch, J.},
  \bibinfo{author}{Glazman, L.~I.} \& \bibinfo{author}{Devoret, M.~H.}
\newblock \bibinfo{title}{Fluxonium: Single cooper-pair circuit free of charge
  offsets}.
\newblock \emph{\bibinfo{journal}{Science}} \textbf{\bibinfo{volume}{326}},
  \bibinfo{pages}{113--116} (\bibinfo{year}{2009}).
\newblock
  \urlprefix\url{https://www.science.org/doi/abs/10.1126/science.1175552}.
\newblock \eprint{https://www.science.org/doi/pdf/10.1126/science.1175552}.

\bibitem{nguyen2019high}
\bibinfo{author}{Nguyen, L.~B.} \emph{et~al.}
\newblock \bibinfo{title}{High-coherence fluxonium qubit}.
\newblock \emph{\bibinfo{journal}{Phys. Rev. X}} \textbf{\bibinfo{volume}{9}},
  \bibinfo{pages}{041041} (\bibinfo{year}{2019}).
\newblock \urlprefix\url{https://link.aps.org/doi/10.1103/PhysRevX.9.041041}.

\bibitem{PhysRevA.86.032324}
\bibinfo{author}{Fowler, A.~G.}, \bibinfo{author}{Mariantoni, M.},
  \bibinfo{author}{Martinis, J.~M.} \& \bibinfo{author}{Cleland, A.~N.}
\newblock \bibinfo{title}{Surface codes: Towards practical large-scale quantum
  computation}.
\newblock \emph{\bibinfo{journal}{Phys. Rev. A}} \textbf{\bibinfo{volume}{86}},
  \bibinfo{pages}{032324} (\bibinfo{year}{2012}).
\newblock \urlprefix\url{https://link.aps.org/doi/10.1103/PhysRevA.86.032324}.

\bibitem{Kreikebaum_2020}
\bibinfo{author}{Kreikebaum, J.~M.}, \bibinfo{author}{O’Brien, K.~P.},
  \bibinfo{author}{Morvan, A.} \& \bibinfo{author}{Siddiqi, I.}
\newblock \bibinfo{title}{Improving wafer-scale josephson junction resistance
  variation in superconducting quantum coherent circuits}.
\newblock \emph{\bibinfo{journal}{Superconductor Science and Technology}}
  \textbf{\bibinfo{volume}{33}}, \bibinfo{pages}{06LT02}
  (\bibinfo{year}{2020}).
\newblock \urlprefix\url{https://dx.doi.org/10.1088/1361-6668/ab8617}.

\bibitem{10.1063/5.0037093}
\bibinfo{author}{Osman, A.} \emph{et~al.}
\newblock \bibinfo{title}{{Simplified Josephson-junction fabrication process
  for reproducibly high-performance superconducting qubits}}.
\newblock \emph{\bibinfo{journal}{Applied Physics Letters}}
  \textbf{\bibinfo{volume}{118}}, \bibinfo{pages}{064002}
  (\bibinfo{year}{2021}).
\newblock \urlprefix\url{https://doi.org/10.1063/5.0037093}.
\newblock
  \eprint{https://pubs.aip.org/aip/apl/article-pdf/doi/10.1063/5.0037093/14545882/064002\_1\_online.pdf}.

\bibitem{doi:10.1126/sciadv.abi6690}
\bibinfo{author}{Zhang, E.~J.} \emph{et~al.}
\newblock \bibinfo{title}{High-performance superconducting quantum processors
  via laser annealing of transmon qubits}.
\newblock \emph{\bibinfo{journal}{Science Advances}}
  \textbf{\bibinfo{volume}{8}}, \bibinfo{pages}{eabi6690}
  (\bibinfo{year}{2022}).
\newblock
  \urlprefix\url{https://www.science.org/doi/abs/10.1126/sciadv.abi6690}.
\newblock \eprint{https://www.science.org/doi/pdf/10.1126/sciadv.abi6690}.

\bibitem{mckay2017efficient}
\bibinfo{author}{McKay, D.~C.}, \bibinfo{author}{Wood, C.~J.},
  \bibinfo{author}{Sheldon, S.}, \bibinfo{author}{Chow, J.~M.} \&
  \bibinfo{author}{Gambetta, J.~M.}
\newblock \bibinfo{title}{Efficient $z$ gates for quantum computing}.
\newblock \emph{\bibinfo{journal}{Phys. Rev. A}} \textbf{\bibinfo{volume}{96}},
  \bibinfo{pages}{022330} (\bibinfo{year}{2017}).
\newblock \urlprefix\url{https://link.aps.org/doi/10.1103/PhysRevA.96.022330}.

\bibitem{han2023multilevel}
\bibinfo{author}{Han, Z.} \emph{et~al.}
\newblock \bibinfo{title}{Multi-level variational spectroscopy using a
  programmable quantum simulator} (\bibinfo{year}{2023}).
\newblock \eprint{2306.02110}.

\bibitem{xu2020high}
\bibinfo{author}{Xu, Y.} \emph{et~al.}
\newblock \bibinfo{title}{High-fidelity, high-scalability two-qubit gate scheme
  for superconducting qubits}.
\newblock \emph{\bibinfo{journal}{Phys. Rev. Lett.}}
  \textbf{\bibinfo{volume}{125}}, \bibinfo{pages}{240503}
  (\bibinfo{year}{2020}).
\newblock
  \urlprefix\url{https://link.aps.org/doi/10.1103/PhysRevLett.125.240503}.

\bibitem{PhysRevLett.125.240502}
\bibinfo{author}{Collodo, M.~C.} \emph{et~al.}
\newblock \bibinfo{title}{Implementation of conditional phase gates based on
  tunable $zz$ interactions}.
\newblock \emph{\bibinfo{journal}{Phys. Rev. Lett.}}
  \textbf{\bibinfo{volume}{125}}, \bibinfo{pages}{240502}
  (\bibinfo{year}{2020}).
\newblock
  \urlprefix\url{https://link.aps.org/doi/10.1103/PhysRevLett.125.240502}.

\bibitem{chu_coupler-assisted_2021}
\bibinfo{author}{Chu, J.} \& \bibinfo{author}{Yan, F.}
\newblock \bibinfo{title}{Coupler-{Assisted} {Controlled}-{Phase} {Gate} with
  {Enhanced} {Adiabaticity}}.
\newblock \emph{\bibinfo{journal}{Physical Review Applied}}
  \textbf{\bibinfo{volume}{16}}, \bibinfo{pages}{054020}
  (\bibinfo{year}{2021}).
\newblock
  \urlprefix\url{https://link.aps.org/doi/10.1103/PhysRevApplied.16.054020}.

\bibitem{PhysRevX.11.021026}
\bibinfo{author}{Ficheux, Q.} \emph{et~al.}
\newblock \bibinfo{title}{Fast logic with slow qubits: Microwave-activated
  controlled-z gate on low-frequency fluxoniums}.
\newblock \emph{\bibinfo{journal}{Phys. Rev. X}} \textbf{\bibinfo{volume}{11}},
  \bibinfo{pages}{021026} (\bibinfo{year}{2021}).
\newblock \urlprefix\url{https://link.aps.org/doi/10.1103/PhysRevX.11.021026}.

\bibitem{PhysRevX.13.031035}
\bibinfo{author}{Ding, L.} \emph{et~al.}
\newblock \bibinfo{title}{High-fidelity, frequency-flexible two-qubit fluxonium
  gates with a transmon coupler}.
\newblock \emph{\bibinfo{journal}{Phys. Rev. X}} \textbf{\bibinfo{volume}{13}},
  \bibinfo{pages}{031035} (\bibinfo{year}{2023}).
\newblock \urlprefix\url{https://link.aps.org/doi/10.1103/PhysRevX.13.031035}.

\bibitem{kono_breaking_2020}
\bibinfo{author}{Kono, S.} \emph{et~al.}
\newblock \bibinfo{title}{Breaking the trade-off between fast control and long
  lifetime of a superconducting qubit}.
\newblock \emph{\bibinfo{journal}{Nature Communications}}
  \textbf{\bibinfo{volume}{11}}, \bibinfo{pages}{3683} (\bibinfo{year}{2020}).
\newblock \urlprefix\url{https://doi.org/10.1038/s41467-020-17511-y}.

\bibitem{acharya_multiplexed_2023}
\bibinfo{author}{Acharya, R.} \emph{et~al.}
\newblock \bibinfo{title}{Multiplexed superconducting qubit control at
  millikelvin temperatures with a low-power cryo-{CMOS} multiplexer}.
\newblock \emph{\bibinfo{journal}{Nature Electronics}}
  \textbf{\bibinfo{volume}{6}}, \bibinfo{pages}{900--909}
  (\bibinfo{year}{2023}).
\newblock \urlprefix\url{https://doi.org/10.1038/s41928-023-01033-8}.

\bibitem{krinner2019engineering}
\bibinfo{author}{{Krinner, S.}} \emph{et~al.}
\newblock \bibinfo{title}{Engineering cryogenic setups for 100-qubit scale
  superconducting circuit systems}.
\newblock \emph{\bibinfo{journal}{EPJ Quantum Technol.}}
  \textbf{\bibinfo{volume}{6}}, \bibinfo{pages}{2} (\bibinfo{year}{2019}).
\newblock \urlprefix\url{https://doi.org/10.1140/epjqt/s40507-019-0072-0}.

\bibitem{shang2013novel}
\bibinfo{author}{Shang, X.}, \bibinfo{author}{Wang, Y.}, \bibinfo{author}{Xia,
  W.} \& \bibinfo{author}{Lancaster, M.~J.}
\newblock \bibinfo{title}{Novel multiplexer topologies based on all-resonator
  structures}.
\newblock \emph{\bibinfo{journal}{IEEE Transactions on Microwave Theory and
  Techniques}} \textbf{\bibinfo{volume}{61}}, \bibinfo{pages}{3838--3845}
  (\bibinfo{year}{2013}).

\bibitem{hong2004microstrip}
\bibinfo{author}{Hong, J.-S.~G.} \& \bibinfo{author}{Lancaster, M.~J.}
\newblock \emph{\bibinfo{title}{Microstrip filters for RF/microwave
  applications}} (\bibinfo{publisher}{John Wiley \& Sons},
  \bibinfo{year}{2004}).

\bibitem{mirhosseini2018superconducting}
\bibinfo{author}{Mirhosseini, M.} \emph{et~al.}
\newblock \bibinfo{title}{Superconducting metamaterials for waveguide quantum
  electrodynamics}.
\newblock \emph{\bibinfo{journal}{Nature communications}}
  \textbf{\bibinfo{volume}{9}}, \bibinfo{pages}{3706} (\bibinfo{year}{2018}).

\end{thebibliography}


\subsection*{Acknowledgements}
\noindent We thank Zhi Hao Jiang, Sitong Tang, Ji Chu, Yanyan Cai, Jiheng Duan, and Peng Zhao for fruitful discussions. This work was supported by the National Natural Science Foundation of China (Grants No.~12322413, No.~92365206).





\end{document}